\documentclass[aps,prd,onecolumn,preprintnumbers,groupedaddress,showpacs,nofootinbib,amssymb]{revtex4-2}
\usepackage{graphicx,color}
\usepackage{amsmath}
\usepackage{amssymb}
\usepackage{amsfonts}
\usepackage{bm}
\usepackage{cancel}
\usepackage{slashed}
\newcommand{\e}{\mathrm{e}}

\allowdisplaybreaks[4] \tolerance=5000

\begin{document}

\title{Is Phantom Divide Crossing in General Relativity Completely Impossible? Shortcomings in canonical and minimally coupled scalar field and
Possible Solutions in $k$-essence Models}
\author{Shin'ichi~Nojiri,$^{1,2}$}
\email{nojiri@nagoya-u.jp}
\affiliation{$^{1)}$ Theory Center, High Energy Accelerator Research Organization (KEK), \\
Oho 1-1, Tsukuba, Ibaraki 305-0801, Japan}
\affiliation{$^{2)}$ Kobayashi-Maskawa Institute for the Origin of Particles and the Universe, Nagoya University, Nagoya 464-8602, Japan}
\author{S.D. Odintsov,$^{3,4,5}$}
\email{odintsov@ieec.cat} \affiliation{$^{3)}$ ICREA, Passeig Luis Companys, 23, 08010 Barcelona, Spain}
\affiliation{$^{4)}$ Institute of Space Sciences (IEEC-CSIC) C. Can Magrans s/n, 08193 Barcelona, Spain}
\affiliation{$^{5)}$ Institut d'Estudis Espacials de Catalunya (IEEC), Edifici RDIT, Campus UPC, 08860 Castelldefels (Barcelona), Spain}
\author{V.K. Oikonomou,$^{6,7}$}
\email{v.k.oikonomou1979@gmail.com;voikonomou@gapps.auth.gr}
\affiliation{$^{6)}$Department of Physics, Aristotle University of Thessaloniki, Thessaloniki 54124, Greece}
\affiliation{$^{7)}$Center for Theoretical Physics, Khazar University, 41 Mehseti Str., Baku, AZ-1096, Azerbaijan}

\begin{abstract}
General relativity has its successes at the local astrophysical
level, however, it seems to be insufficient in describing the
Universe at large scales. In this work, we investigate how the
generalized scalar field theories in the context of general
relativity can accommodate a phantom-to-quintessence transition,
which may be an essential element of realistic Dark Energy
scenarios in the late Universe. As we demonstrate in a very
detailed manner, this is impossible for a canonical and minimally
coupled single scalar field theory, but it may be possible for
$k$-essence theories. We point out how the ghost instabilities may
be eliminated, and we analyze the quantitative features of a
$k$-essence theory that may realize a phantom-to-quintessence
transition in the late Universe. We also qualitatively compare the
difficulties and fine-tunings required for $k$-essence theories to
realize a phantom-to-quintessence transition, and how such a
transition is naturally realized in modified gravity, without
unnecessary fine-tunings and ghost eliminations.
\end{abstract}

\maketitle

\section{Introduction}

A Universe filled with a perfect fluid, like dark energy, is
characterized by the equation of state (EoS) parameter $w$, which
is the ratio of the pressure $p$ and the energy density $\rho$ of
the fluid, $w=\frac{p}{\rho}$. The dark energy with $w<-1$ is
called phantom. Recent Dark Energy Spectroscopic Instrument (DESI)
observations~\cite{DESI:2024mwx, DESI:2025zgx} indicate that there
might have been a transition from a phantom regime to a
quintessential regime, which is called the phantom-to-quintessence
crossing \cite{Cortes:2024lgw, Colgain:2024xqj, Giare:2024smz,
Shlivko:2024llw}, although the original meaning of the phantom
crossing is the transition from the non-phantom Universe to the
phantom Universe. We should note, however, the dark-energy
equation of state is not directly observed but it is inferred
within a chosen parametrization or physical model. The apparent
phantom crossing can depend strongly on the assumed model space.
In addition, the results of DESI do not automatically imply that a
physical scalar-field dark-energy component crosses the phantom
divide. There could be three possibilities: 1. a crossing in a
parameterized equation of state, 2. an effective crossing
generated by interactions or modified gravity, and 3. a true
crossing of the physical dark-energy component. We also consider
the first possibility of a crossing in a parameterized equation of
state. The second possibility of interacting dark-sector models
has been considered in \cite{Petri:2025swg, Chakraborty:2025syu,
Shah:2024rme}, where it has been shown that an interacting or
coupled fluid model can be degenerate with an evolving dark-energy
parametrization at the background level while replacing the
apparent phantom crossing by a sign change in the dark-sector
interaction. Another example is the scenario described in Ref.
\cite{Dinda:2025iaq}, where thawing quintessence with non-zero
curvature can fit recent data without requiring a physical
crossing of the phantom divide.

Since the DESI data were available, and also previously, a large
stream of articles has been produced towards explaining the data
accurately and also pointing towards resolving tensions in modern
cosmology, see for example Refs.~\cite{Odintsov:2024woi,
Dai:2020rfo, He:2020zns, Nakai:2020oit, DiValentino:2020naf,
Agrawal:2019dlm,Ye:2020btb,Desmond:2019ygn, Hogas:2023pjz,
OColgain:2018czj, Krishnan:2020obg, Colgain:2019joh, Lee:2022cyh,
Krishnan:2021dyb, Ye:2021iwa, Ye:2022afu, Verde:2019ivm,
Menci:2024rbq, Adil:2023ara, Reeves:2022aoi, Ferlito:2022mok,
Vagnozzi:2021quy, DiValentino:2020evt, Sabogal:2024yha,
CosmoVerseNetwork:2025alb, Odintsov:2025kyw, Odintsov:2025jfq,
Kessler:2025kju, Nojiri:2025low, Nojiri:2025uew,
Odintsov:2026fyc,Han:2024qbw,You:2024hit,He:2024hmc,Han:2024qbw,You:2024hit,He:2024hmc,Tang:2024gtq,Wolf:2025acj,Wolf:2025jed,Wolf:2024stt,Li:2026xaz,Li:2025owk}
and references therein. We often call the phantom crossing
observed by the DESI an inverse phantom crossing. Because it is
known that the phantom dark energy cannot be realized by the
canonical scalar field, whose action is given by,
\begin{align}
\label{cnnclsclr}
S_\eta = \int d^4 x \sqrt{-g} \left( - \frac{1}{2} \partial^\mu \eta \partial_\mu \eta - V(\eta) \right) \, ,
\end{align}
see for example Ref.~\cite{Elizalde:2004mq, Nojiri:2005sr,
Vikman:2004dc}. Here $\eta$ is the canonical scalar field and
$V(\eta)$ is the potential of $\eta$. Now, it is known that the
description of General Relativity (GR) of inflation and dark
energy is comprised of single scalar field theories and
$k$-essence theories. In this work, we aim to investigate whether
phantom-to-quintessence transitions can be done in the context of
GR. Although for modified gravity \cite{Nojiri:2017ncd,
Capozziello:2011et, Faraoni:2010pgm, Nojiri:2006ri,
Nojiri:2010wj}, this is a matter of simple modelling, without
falling into ghost instabilities and inconsistencies, in the
context of GR, a phantom-to-quintessence transition is not a
simple thing to achieve. We shall concretely prove that a
phantom-to-quintessence transition is impossible in the context of
single scalar field theory. However, this can be achieved in
$k$-essence theories \cite{Chiba:1999ka, Armendariz-Picon:2000nqq,
Armendariz-Picon:2000ulo}, with some fine-tuning. The $k$-essence
model might have ghosts, so we will use the systematic analysis of
the reconstruction and the stability in the $k$-essence model made
in Ref.~\cite{Matsumoto:2010uv}.

\section{General Considerations of Phantom Crossing in Single Scalar Field GR Framework}

Let us discuss in general whether phantom to quintessence are allowed in the context of GR materialized by minimally coupled single scalar field theories or even extensions like $k$-essence theories, which still fall in the GR context.
The possibility that the dark energy EoS parameter $w \equiv p/\rho$ may evolve dynamically and possibly cross the phantom divide line $w=-1$ has attracted interest in the literature in the past, see for example, \cite{Elizalde:2004mq, Nojiri:2005sr, Vikman:2004dc}.
Recently, this phantom crossing issue has become popular due to the new DESI data that indicate a dynamical dark energy evolving from a phantom EoS parameter to a quintessential EoS parameter.
Observationally, such a crossing is not excluded, while theoretically it provides a sharp discriminator between the various different classes of dark energy models, spanning from modified gravity to GR-compatible models such as $k$-essence and single scalar field theories.
In this section, we shall provide a unified and detailed discussion of whether a phantom to quintessence transition is possible within the context of GR, when the dark sector is modelled by a single scalar field.
We shall assume a flat background metric and specifically, a Friedmann-Lema\^{i}tre-Robertson-Walker (FLRW) Universe,
\begin{align}
\label{metricflrw}
ds^2 = - dt^2 + a(t)^2 \sum_{i=1,2,3} \left(dx^i\right)^2\, ,
\end{align}
The Einstein equations for the canonical scalar field theory are
\begin{align}
G_{\mu\nu} = {M_\mathrm{Pl}}^{-2} T_{\mu\nu}\, ,
\end{align}
with $M_\mathrm{Pl}^2 = (8\pi G)^{-1}$.
For a homogeneous scalar field $\phi(t)$, the Friedmann equations read
\begin{align}
\label{Friedmann}
H^2 =&\, \frac{1}{3{M_\mathrm{Pl}}^2}\rho\, , \nonumber \\
\dot H =&\, -\frac{1}{2{M_\mathrm{Pl}}^2}(\rho+p)\, .
\end{align}
The equation-of-state parameter is
\begin{align}
w = \frac{p}{\rho}\, .
\end{align}
As long as $\rho>0$, a  phantom regime corresponds to $w<-1$, or
equivalently expressed, $\rho+p<0$, which implies $\dot H>0$. If
$\rho$ is negative, the first equation in \eqref{Friedmann} has no
solution, that is, there is no expansion of the universe. For this
reason, we assume that $\rho$ is not negative. For the minimally
coupled canonical scalar field, we have,
\begin{align}
\mathcal{L} = -X - V(\phi)\, , \quad X \equiv \frac12
\partial_\mu\phi\partial^\mu\phi\, .
\end{align}
For a homogeneous field,
\begin{align}
\rho = -X + V\, , \quad p =  -X - V\, .
\end{align}
Therefore,
\begin{align}
\rho + p =  -2X = \dot\phi^2 \ge 0\, .
\end{align}
This identity is independent of the sign or form of the scalar field potential $V(\phi)$.
Consequently, we have,
\begin{align}
w = \frac{ -X-V}{ -X+V} \ge -1\, ,
\end{align}
with equality only if $\dot\phi=0$ or $\dot{\phi}\ll 1$, which is
basically similar to the slow-roll condition. Now, let us state
flat out that simply, a canonical scalar field in GR cannot
realize a phantom phase, nor can it cross $w=-1$, without invoking
additional ghost scalar fields. This result is purely structural
due to the construction of the theory and does not depend on the
choice of potential or the initial conditions. Note also that even
if we have negative potentials, this result holds true in general.
Negative potentials are motivated by various UV-completions of GR,
such as AdS vacua, moduli stabilization, or ekpyrotic cosmology,
see also \cite{Heard:2002dr}. However, for late-time dark energy
physics, such negative potentials often lead to inconsistencies.
But even allowing these negative potentials $V<0$, the identity
$\rho+p=\dot\phi^2$ remains valid. Phantom behavior looks to be
still impossible. In addition, negative potentials may lead to
loss of accelerated expansion, or even vanishing total energy
density $\rho\to 0$, and possibly divergences in the EoS
parameter, when $\dot{\phi}^2=V$ if there is not additional
contribution. These are strong finite-time singularities which
cannot be resolved or ignored in the theory, see for example
\cite{deHaro:2023lbq, Nojiri:2005sx, Bahamonde:2016wmz,
Odintsov:2016plw}.
%Thus, the dark energy evolution may become complicated or lose its consistency, so let us ignore the negative potentials.
In this paper, we ignore the negative potentials so that the dark
energy evolution does not become complicated or lose its
consistency. We should note, however, that as shown in
\cite{Mukherjee:2025myk}, a negative cosmological constant with
dynamical dark energy can allow viable quintessence-like regions
in the dark-energy parameter space.

Let us also consider the $k$-essence case, so the most general single-field scalar theory with second-order equations of motion has the following Lagrangian,
\begin{align}
\mathcal{L} = P(\phi, X)\, .
\end{align}
The energy density and pressure are
\begin{align}
\rho = 2XP_X - P\, , \quad p = P\, .
\end{align}
Therefore,
\begin{align}
\rho+p = 2X P_X\, .
\end{align}
A phantom phase requires $P_X < 0$, and the crossing $w=-1$ would
require $P_X = 0$. In Ref.~\cite{Elizalde:2004mq, Nojiri:2005sr,
Vikman:2004dc}, it was demonstrated that in single-field
$k$-essence models, the condition $P_X=0$ corresponds to a
vanishing kinetic term for the cosmic perturbations, and at that
point, the sound speed would become ill-defined. In addition, the
very own Cauchy problem would become ill-defined, thus the theory
would become strongly coupled or even ghostly unstable. Hence,
although GR itself does not forbid the phantom divide line $w=-1$
crossing, the realization of such crossing cannot be done in the
context of a single scalar field theory and its $k$-essence
extensions, although we shall provide some fine-tuned examples of
$k$-essence phantom transitions. On the contrary, in multi-scalar
field theories and in modified gravity theories, the
phantom-divide line transition can easily occur without violating
fundamental principles.

Let us be more quantitative at this point and provide the no-go features of phantom-to-quintessence transitions in single scalar field theory, and how this can be realized by a fine-tuned $k$-essence theory.
So consider the single scalar field Lagrangian
\begin{align}
\mathcal{L} = -\frac{1}{2}\,\partial_\mu \phi\,\partial^\mu \phi - V(\phi), \qquad V(\phi) > 0\, ,
\end{align}
in a spatially flat FLRW background.
For a canonical scalar field, we have,
\begin{align}
\rho_\phi = \frac{1}{2}\dot{\phi}^2 + V(\phi)\, , \quad p_\phi = \frac{1}{2}\dot{\phi}^2 - V(\phi)\, ,
\end{align}
so the EoS parameter is
\begin{align}
w_\phi = \frac{p_\phi}{\rho_\phi} = \frac{\tfrac{1}{2}\dot{\phi}^2 - V(\phi)} {\tfrac{1}{2}\dot{\phi}^2 + V(\phi)}\, .
\end{align}
Since $\dot{\phi}^2 \ge 0$ and $V(\phi) > 0$, we always have
\begin{align}
w_\phi \ge -1\, ,
\end{align}
with the equality holding true only when $\dot{\phi}=0$ or $\dot{\phi}\ll 1$ which is basically the realization of the slow-roll condition.
On the other hand, the phantom regime
requires
\begin{align}
w < -1\, ,
\end{align}
which is impossible to achieve for a canonical scalar field.
Therefore, no phantom-to-quintessence transition is possible because the theory never enters the phantom regime in the first place.
The phantom crossing would require,
\begin{align}
\dot{\phi}^2 < 0\, ,
\end{align}
which violates the canonical kinetic structure.

Let us further elaborate and assume that $\dot{\phi}^2 = \beta(\phi)$, where $\beta(\phi) \ge 0$ by definition.
The EoS parameter becomes,
\begin{align}
\label{beta1}
w_\phi(\phi) = \frac{\frac{\beta(\phi)}{2} - V(\phi)}{\frac{\beta(\phi)}{2} + V(\phi)}\, ,
\end{align}
or equivalently,
\begin{align}
\label{beta2}
w_\phi + 1 = \frac{\beta(\phi)}{\tfrac{1}{2}\beta(\phi) + V(\phi)} \ge 0\, .
\end{align}
So we have $w=-1$ only if $\beta(\phi)=0$ or $\beta(\phi)\ll 1$, and $w>-1$ whenever $\beta(\phi)>0$, but no choice of $\beta(\phi)$ allows $w<-1$.
Even allowing $\beta(\phi)$ to vary arbitrarily or discontinuously does not change the result, the sign constraint is fundamental.
Hence, the kinetic function $\dot{\phi}^2=\beta(\phi)$ cannot produce any phantom behavior or a phantom-to-quintessence transition in a canonical scalar theory.

We may consider a generalized $\phi$-dependent constant-roll-like
behavior for which $\dot{\phi}^2 = \beta(\phi) V(\phi)$, which
includes the slow-roll and constant-roll evolution as special
cases. If we redefine $\beta(\phi) V(\phi) \to \beta(\phi)$, by
the arguments after \eqref{beta1},
%Consider now a generalized $\phi$-dependent constant-roll-like behavior for which $\dot{\phi}^2 = \beta(\phi) V(\phi)$, which includes the slow-roll and constant-roll evolution as special cases.
%The energy density and pressure take the form,
%\begin{align}
%\rho_\phi =&\, \frac{1}{2}\beta(\phi)V(\phi) + V(\phi) = V(\phi)\left(1 + \frac{1}{2}\beta(\phi)\right)\, , \\
%p_\phi =&\, \frac{1}{2}\beta(\phi)V(\phi) - V(\phi) = V(\phi)\left(\frac{1}{2}\beta(\phi) - 1\right)\, .
%\end{align}
%Thus, we have,
%\begin{align}
%w_\phi = \frac{\tfrac{1}{2}\beta(\phi) - 1}{\tfrac{1}{2}\beta(\phi) + 1}\, ,
%\end{align}
%so,  $w=-1$ if $\beta(\phi)=0$ or $\beta(\phi)\ll 1$, and $w>-1$ if $\beta(\phi)>0$.
%In order to have $w<-1$, we would require the condition $\beta(\phi)<0$, but this would imply,
%\begin{align}
%\dot{\phi}^2 < 0\, ,
%\end{align}
%which is unphysical.
we find that the model could be unphysical. Hence, in order to
have a phantom-to-quintessence transition, one must abandon
single-field canonical and minimally coupled scalar theory, and
should invoke ghost fields or $k$-essence, a multi-scalar theory,
or even non-Lorentz-invariant scalar dynamics.

About the $k$-essence possibility, some models can in principle achieve the phantom-to-quintessence transition, for example,
\begin{align}
P(X) = -\,X + \frac{X^{2}}{M^{4}}\, ,
\end{align}
where $M$ is a mass scale.
This theory admits a background solution satisfying the condition,
\begin{align}
P_{X} = 0
\end{align}
at a stationary point of the kinetic function. Although this
allows for a phantom-to-quintessence transition, it may require
fine-tuning of the background solution, and it may have marginal
stability, and might require the necessity of controlling
higher-derivative operators in the effective theory, thus
fine-tuning. Hence, phantom crossing in such $k$-essence models
is, in principle, possible but highly non-generic and
theoretically marginally consistent. Nevertheless, it is possible
for $k$-essence models to achieve this phantom-to-quintessence
transition. In the next sections, we further analyze this
possibility.

\section{Essential Features of a Class of $k$-essence Models}

In this section, we review the arguments about the reconstruction and the stability in the $k$-essence model developed in Ref.~\cite{Matsumoto:2010uv}.
The reconstruction is a formulation to construct a model that gives a solution realizing a given evolution of the Universe.
Especially, we consider the reconstruction of the model, realizing the evolution of the accelerating expansion of the Universe.
We also review the conditions for the stability of the given solution in the framework of the model.

\subsection{Pure Kinetic Model}

\subsubsection{Reconstruction of model}

We first consider the model where the action for a scalar field $\phi$ is only given by the kinetic term, which was the original $k$-essence model, as follows,
\begin{align}
\label{M1} S = \int d^4x \sqrt{-g}\left\{\frac{R}{2\kappa^2}
 -K \left( q\left(\phi\right)\partial^\mu \phi \partial _\mu \phi \right) \right\}\, ,
\end{align}
where $K$ and $q$ are adequate functions of $q\left(\phi\right)\partial^\mu \phi \partial _\mu \phi$ and $\phi$, respectively.
We should note that as long as the signature of $q\left(\phi\right)$ is not changed, $q\left(\phi\right)$ can be absorbed into the redefinition of $\phi$ as follows,
\begin{align}
\label{varphi}
\varphi = \int d\phi \sqrt{ \left| q\left(\phi\right) \right|}\, ,
\end{align}
and we obtain,
\begin{align}
\label{vph}
q\left(\phi\right)\partial^\mu \phi \partial _\mu \phi = \mathrm{sign} \left( q \right) \partial^\mu \varphi \partial_\mu \varphi \, .
\end{align}
Here $\mathrm{sign} \left( q \right)$ is defined by,
\begin{align}
\label{sgn}
\mathrm{sign} \left( q \right) \equiv \left\{
\begin{array}{ll}
1 & \mbox{when}\ q>0 \\
 -1 & \mbox{when}\ q< 0
\end{array} \right. \, .
\end{align}
We consider the formulation of the reconstruction in this model \eqref{M1}.

In the background of the FLRW Universe, it is possible to assume the scalar field $\phi$ depends only on the time coordinate $t$.
This is because the redefinition of $\phi$ to a new scalar field $\tilde\phi$, $\phi=\phi\left(\tilde\phi\right)$, can be absorbed into $q\left(\phi\right)$, as follows,
\begin{align}
\label{vph2} q\left(\phi\right)\partial^\mu \phi \partial _\mu \phi = \tilde q\left(\tilde\phi\right)\partial^\mu \tilde \phi \partial _\mu \tilde \phi \, , \quad
\tilde q\left(\tilde\phi\right) \equiv q\left(\phi\left(\tilde\phi\right)\right) \phi'\left(\tilde\phi\right)^2\, .
\end{align}
so we may identify the scalar field $\phi$ with the time coordinate $t$.
Then, we obtain the following first FLRW equation and the equation given by the variation of $\phi$, as follows,
\begin{align}
\label{M3}
\frac{3}{\kappa^2}H^2 =&\, K \left(-q \left(t\right) \right) + 2K' \left(-q\left(t\right)\right)q\left(t\right)\, , \\
\label{M4}
0 =&\, 2K'' \left( -q\left(t\right)\right)q\left(t\right)q'\left(t\right) - \left( 6H q\left(t\right) + q'\left(t\right) \right)K' \left(-q\left(t\right) \right)\, .
\end{align}
Here $H$ is the Hubble rate defined by $H = \dot a \left(t\right)/a\left(t\right)$.
Eq.~\eqref{M4} can be integrated to give,
\begin{align}
\label{M5}
a\left(t\right)^6 = \frac{1}{q\left(t\right)} \left(\frac{K_0}{K' \left( -q\left(t\right) \right)}\right)^2\, .
\end{align}
By differentiating \eqref{M3} with respect to $t$ and using \eqref{M5}, we obtain,
\begin{align}
\label{M6}
\frac{6}{\kappa^2} \dot H = -6 K_0 a\left(t\right)^{-3} \sqrt{q\left(t\right)}\, ,
\end{align}
which gives,
\begin{align}
\label{M7}
q\left(t\right) = \frac{a\left(t\right)^6 (H' \left(t\right))^2}{\kappa^4 K_0^2}\, .
\end{align}
Eq.~\eqref{M7} gives the explicit form of $q\left(t\right)$.
In principle, we can solve \eqref{M7} with respect to $t$ by using an adequate function $f$ as $t = f(q)$.
By using the explicit form of $f(q)$, we find that Eq.~\eqref{M5} gives a form of $K'(-q)$ as follows,
\begin{align}
\label{M9}
K' (-q) = \frac{K_0}{\sqrt{q}} a\left(f \left(q\right) \right)^{-3}\, .
\end{align}
For an arbitrary time-development of $a$ or $H$ given by $a = a\left(t\right)$, Eqs.~\eqref{M7} and \eqref{M9} give the forms of the functions $K$ and $q$ realising the given time-development.

We should note, however, $\dot H $ cannot change its sign as we can see from \eqref{M6}.
Then the transition between the non-phantom phase ($ \dot H < 0$ ) and the phantom phase ( $ \dot H > 0$ ) does not occur.
We can extend the model \eqref{M1} to include the matter fluids with constant EoS parameters $w_i$.
Because the energy density of the matter is given by $\sum _i \rho_{0i}a^{-3\left( 1+w_i \right)}$ with constants $\rho_{0i}$, the FLRW equation \eqref{M3} is modified as follows,
\begin{align}
\label{M15}
\frac{3}{\kappa^2}H^2 = K(-q\left(t\right)) + 2K'(-q\left(t\right))q\left(t\right) + \sum _i \rho_{0i}a^{-3\left(1+w_i\right)}\, .
\end{align}
Eq.~\eqref{M4} is not changed and we reobtain \eqref{M5} although Eq.~\eqref{M6} is modified to be,
\begin{align}
\label{M16}
\frac {6}{\kappa^2} \dot H = -6 K_0 a\left(t\right)^{-3} \sqrt{q\left(t\right)} - \sum _i 3\left(1+w_i\right) \rho _{0i} a^{-3\left(1+w_i\right)}\, ,
\end{align}
and we obtain,
\begin{align}
\label{M17}
q\left(t\right) = \frac {a\left(t\right)^6 \left( H'\left(t\right) + \frac{\kappa^2}{2} \sum _i \left(1+w_i\right) \rho _{0i} a^{-3\left(1+w_i\right)} \right)^2}{\kappa^4 K_0^2}\, ,
\end{align}
which gives the explicit form of $q\left(t\right)$ as in \eqref{M7}.
If Eq.~\eqref{M17} can be solved with respect to $t$ by using an adequate function $f$ as $t = f(q)$, again, Eq.~\eqref{M5} gives an explicit form of $K' (-q)$ by \eqref{M9}.
Some examples were given in Ref.~\cite{Matsumoto:2010uv}.
In the case of the $\Lambda$CDM model, $K$ becomes a constant corresponding to the cosmological constant.

\subsubsection{Stability of solutions}

We now investigate the stability of the solution.
We start with the first and second FLRW equations and the field equation obtained from Eq.~\eqref{M1},
\begin{align}
\label{M2_1}
\frac {3}{\kappa^2}H^2 =&\, K \left( -q \dot \phi^2 \right) + 2K' \left( -q \dot \phi^2 \right)q \dot \phi^2\, , \quad
\frac {1}{2 \kappa^2} \left( 3H^2 + 2 \dot H \right) = \frac {1}{2}K \left( -q \dot \phi^2 \right)\, , \\
\label{M2_3}
0=&\, \frac {d}{dt} \left( 2a^3 q^\frac{1}{2} \dot \phi K'\left(-q \dot \phi^2\right) \right) \, .
\end{align}
The FLRW equations in \eqref{M2_1} indicate that,
\begin{align}
\label{M2_4}
K' (-Q) Q = - \frac {\dot H}{\kappa^2}\, ,
\end{align}
where $Q \equiv q {\dot \phi}^2$.
Substituting Eq.~\eqref{M2_4} into Eq.~\eqref{M2_3}, we obtain $\frac{d}{dt} \left(2a^3 Q^{-\frac {1}{2}} \frac {\dot H}{\kappa^2} \right) = 0$, which can be integrated and gives,
\begin{align}
\label{M2_6}
\frac{1}{\kappa^2}a^3 Q^{- \frac {1}{2}} \dot H \equiv K_0\, ,
\end{align}
where $K_0$ is a constant of the integration. We now write the
form of $a\left(t\right)$ as $a\left(t\right) = a_0
\e^{h\left(t\right)}$ and consider the perturbation of the
function $h\left(t\right)$: $h\left(t\right) = h_0\left(t\right) +
\delta h \left(t\right)$ from the background solution
$h_0\left(t\right)$. By using \eqref{M2_6}, we find the variation
of $Q$ as follows,
\begin{align}
\label{dQ}
\delta Q = \frac{{a_0}^6 \e^{6h_0}}{\kappa ^4 K_0 ^2} \left( 6 \delta h {\ddot h_0}^2 +2 \ddot h_0 \delta \ddot h \right)
= Q_0 \left( 6 \delta h + \frac{2}{\ddot h_0} \delta \ddot h \right) \, .
\end{align}
In the background, we have $H=\dot h_0$, and we write $Q=Q_0=\frac{{a_0}^6 \e^{6h_0}{\ddot h_0}^2}{\kappa ^4 K_0 ^2}$
By differentiating \eqref{M2_4} with respect $t$, we obtain,
\begin{align}
\label{M2_4dff}
 - K''(-Q)Q\dot Q + K'(-Q) \dot Q = - K''(-Q)Q\dot Q - \frac{\dot Q \dot H}{Q\kappa^2}= - \frac{\ddot H}{\kappa^2} \, ,
\end{align}
On the other hand, by differentiating \eqref{M2_6}, we find
\begin{align}
\label{M2_6dff}
\dot Q = Q \left( 6H + \frac{2 \ddot H}{\dot H} \right) \, .
\end{align}
By deleting $\dot Q$ in \eqref{M2_4dff} by using \eqref{M2_6dff}, we obtain
\begin{align}
\label{JM1}
K'' (-Q) Q^2 = - \frac{\dot H}{\kappa^2} + \frac{\ddot H}{\kappa^2\left( 6H + \frac{2 \ddot H}{\dot H} \right)} \, .
\end{align}
In the background, Eq.~\eqref{JM1} gives
\begin{align}
\label{Kdd}
K'' (-Q_0)Q_0 ^2 = - \frac{\ddot h_0}{\kappa ^2} + \frac {\dddot h_0}{\kappa ^2 \left( 6 \dot h_0 + \frac{2 \dddot h_0}{\ddot h_0} \right)}\, .
\end{align}
By considering the variation $h\left(t\right) = h_0\left(t\right) + \delta h \left(t\right)$, the variation of the first equation in \eqref{M2_1}, that is,
$\frac {3}{\kappa^2}H^2 = K \left( -Q \right) + 2K' \left( -Q \right) Q$ gives
\begin{align}
\label{M2_1dlt}
\frac {6}{\kappa ^2} \dot h_0 \delta \dot h = \left( K' (-Q_0)-2K'' (-Q_0)Q_0 \right) \delta Q \, .
\end{align}
By using \eqref{M2_4} in the background $K' (-Q_0) Q_0 = - \frac {\ddot h_0}{\kappa^2}$, \eqref{dQ}, and \eqref{Kdd},
%Then
we obtain a rather simple differential equation for $\delta
h\left(t\right)$,
\begin{align}
\label{M2_9}
\delta \ddot h -  \frac {3 \dot h_0 \ddot h_0 + \dddot h_0}{\ddot h_0}  \delta \dot h + 3 \ddot h_0 \delta h = 0\, .
\end{align}
Because Eq.~\eqref{M2_9} only contains $h_0$ and its derivatives, if $h_0$ is given, even if we do not know the explicit forms of $q$ and $K$, we can find the stability of the solution.
By defining $x\equiv \delta h$ and $y= \delta\dot h$, we rewrite Eq.~\eqref{M2_9} as follows,
\begin{align}
\label{N3}
\frac{d}{dt} \left(\begin{array}{c} x \\ y \end{array} \right) = \left( \begin{array}{cc} 0 & 1 \\ - 3 \ddot h_0 & \frac {3 \dot h_0 \ddot h_0 + \dddot h_0}{\ddot h_0}
\end{array} \right)
\left(\begin{array}{c} x \\ y \end{array} \right) \, .
\end{align}
For the solution to be stable, the real part of all the eigenvalues of the 2$\times$2 matrix in \eqref{N3} should be negative:
\begin{align}
\label{N4}
\Re \lambda_\pm < 0\, ,\quad \lambda_\pm \equiv \frac{1}{2}\left[ \frac {3 \dot h_0 \ddot h_0
+ \dddot h_0}{\ddot h_0} \pm \sqrt{ \left( \frac {3 \dot h_0 \ddot h_0 + \dddot h_0}{\ddot h_0} \right)^2 - 12 \ddot h_0} \right] \, ,
\end{align}
which requires $\ddot h_0>0$.

We may investigate the stability when matter fluids are present, as in \eqref{M15}.
The equations corresponding to \eqref{M2_1} and \eqref{M2_3} are given by,
\begin{align}
\label{M3_1}
\frac{3}{\kappa^2}H^2 =&\, K(-Q) + 2K'(-Q)Q + \sum _i \rho _{0i} a^{-3\left(1+w_i\right)}\, ,\\
\label{M3_2}
\frac{1}{2 \kappa^2} \left( 3H^2 + 2 \dot H \right) =&\,  \frac{1}{2} K(-Q) - \frac{1}{2} \sum _i w_i \rho _{0i} a^{-3\left(1+w_i\right)}\, ,\\
\label{M3_3}
0=&\, \frac{d}{dt} \left( 2a^3Q^{\frac{1}{2}}K'(-Q) \right)\, .
\end{align}
By combining \eqref{M3_1} and \eqref{M3_2}, instead of
\eqref{M2_4}, we obtain
\begin{align}
\label{MN1}
\frac{\dot H}{\kappa^2} =&\, K'(-Q)Q  - \frac{1}{2} \sum _i \left(1+w_i\right) \rho _{0i} a^{-3\left(1+w_i\right)}\, .
\end{align}
By deleting $K'(-Q)$ in \eqref{M3_3} using \eqref{MN1}, instead of \eqref{M2_6}, we obtain
\begin{align}
\label{M2_6matter}
a^3 Q^{- \frac {1}{2}} \left( \frac{\dot H}{\kappa^2} + \frac{1}{2} \sum _i \left(1+w_i\right) \rho _{0i} a^{-3\left(1+w_i\right)} \right)= K_0\, ,
\end{align}
%Eq.~\eqref{M3_3} gives Eq.~\eqref{M2_6} and therefore Eqs.~\eqref{dQ} and \eqref{Kdd}, again.
Under the perturbation $h \rightarrow h_0 + \delta h$, Eq.~\eqref{M2_6matter} gives
%Eqs.~\eqref{dQ} and \eqref{Kdd} are not changed, but
\begin{align}
\label{M2_6matterB}
\delta Q=&\, 6Q_0 \delta h
+ \frac{2Q_0 \left( \frac{\delta h''}{\kappa^2} - \frac{3}{2} \sum _i \left(1+w_i\right)^2 \rho _{0i} {a_0}^{-3\left(1+w_i\right)} \e^{-3\left(1+w_i\right)h_0} \delta h \right)}
{\left( \frac{{h_0}''}{\kappa^2} + \frac{1}{2} \sum _i \left(1+w_i\right) \rho _{0i} {a_0}^{-3\left(1+w_i\right)} \e^{-3\left(1+w_i\right)h_0} \right)}
\, ,
\end{align}
Differentiating \eqref{MN1} and \eqref{M2_6matter} with respect $t$, we obtain
\begin{align}
\label{MN1BB}
\frac{\ddot H}{\kappa^2} =&\, \left( - K''(-Q)Q + K'(-Q)\right) \dot Q  + \frac{3}{2} \sum _i \left(1+w_i\right)^2 \rho _{0i} a^{-3\left(1+w_i\right)} H\, , \\
\label{M2_6matterBB}
\dot Q=&\, 6Q H
+ \frac{2Q \left( \frac{\ddot H}{\kappa^2} - \frac{3}{2} \sum _i \left(1+w_i\right)^2 \rho _{0i} a^{-3\left(1+w_i\right)} \dot H \right)}
{\left( \frac{\dot H}{\kappa^2} + \frac{1}{2} \sum _i \left(1+w_i\right) \rho _{0i} a^{-3\left(1+w_i\right)} \right)} \, .
\end{align}
By combining \eqref{MN1BB} and \eqref{M2_6matterBB}, we can delete $\dot Q$.
Furthermore, by using \eqref{MN1} in the obtained expression, we can express $K''(-Q)Q^2$ in terms of $h$ and the derivatives of $h$,
\begin{align}
\label{MN1CC}
\frac{\ddot H}{\kappa^2} =&\, \left( - K''(-Q)Q^2 + \frac{\dot H}{\kappa^2}
+ \frac{1}{2} \sum _i \left(1+w_i\right) \rho _{0i} a^{-3\left(1+w_i\right)} \right) \left\{ 6 H
+ \frac{2 \left( \frac{\ddot H}{\kappa^2} - \frac{3}{2} \sum _i \left(1+w_i\right)^2 \rho _{0i} a^{-3\left(1+w_i\right)} \dot H \right)}
{\left( \frac{\dot H}{\kappa^2} + \frac{1}{2} \sum _i \left(1+w_i\right) \rho _{0i} a^{-3\left(1+w_i\right)} \right)} \right\} \nonumber \\
&\, + \frac{3}{2} \sum _i \left(1+w_i\right)^2 \rho _{0i} a^{-3\left(1+w_i\right)} H\, , \nonumber \\
K''(-Q)Q^2 =&\, \frac{\dot H}{\kappa^2}
+ \frac{1}{2} \sum _i \left(1+w_i\right) \rho _{0i} a^{-3\left(1+w_i\right)}
+ \frac{ - \frac{\ddot H}{\kappa^2} + \frac{3}{2} \sum _i \left(1+w_i\right)^2 \rho _{0i} a^{-3\left(1+w_i\right)} H}
{6 H
+ \frac{2 \left( \frac{\ddot H}{\kappa^2} - \frac{3}{2} \sum _i \left(1+w_i\right)^2 \rho _{0i} a^{-3\left(1+w_i\right)} \dot H \right)}
{\left( \frac{\dot H}{\kappa^2} + \frac{1}{2} \sum _i \left(1+w_i\right) \rho _{0i} a^{-3\left(1+w_i\right)} \right)}} \, .
\end{align}
On the other hand, Eq.~\eqref{M2_1dlt} is modified by the contributions from matter,
\begin{align}
\label{M2_1dlt_matter}
\frac {6}{\kappa ^2} \dot h_0 \delta \dot h = \left( K' (-Q_0)-2K'' (-Q_0)Q_0 \right) \delta Q
 - 3 \sum _i \left(1+w_i\right) \rho _{0i} {a_0}^{-3\left(1+w_i\right)} \e^{-3\left(1+w_i\right)h_0} \delta h \, .
\end{align}
Deleting $\delta Q$ by using \eqref{M2_6matterB}, $K' (-Q_0)Q_0$
by using \eqref{MN1} and $K'' (-Q_0){Q_0}^2$ by using
\eqref{MN1CC}, we obtain,
\begin{align}
\label{M3_4}
0 =&\, \delta \ddot h -b \delta \dot h -c \delta h\, ,\\
\label{M3_5}
b \equiv &\, 3 \frac{\frac{1}{\kappa^2}(3 \dot h_0 \ddot h_0 + \dddot h_0)
 - \frac{3}{2}\sum _i w_i \left(1+w_i\right) \rho _{0i} a_0^{-3\left(1+w_i\right)}\e^{-3\left(1+w_i\right)h_0}\dot h_0}
 {\frac {3}{\kappa^2} \ddot h_0 + \frac{3}{2} \sum _i \left(1+w_i\right) \rho _{0i} a_0^{-3\left(1+w_i\right)}\e^{-3\left(1+w_i\right)h_0}}\, ,\\
\label{M3_6}
c \equiv &\, -\frac{\kappa^2}{2}\frac{\frac{18}{\kappa^4}\ddot h_0^2 -3 \sum_i \left(1+w_i\right) \rho _{0i}a_0^{-3\left(1+w_i\right)}
\e^{-3\left(1+w_i\right)h_0}\frac{1}{\kappa^2} \left(6w_i \ddot h_0 + \left(1+w_i\right)\frac{\dddot h_0}{\dot h_0} \right)}
{\frac{3}{\kappa^2} \ddot h_0 + \frac{3}{2} \sum _i \left(1+w_i\right) \rho _{0i} a_0^{-3\left(1+w_i\right)} \e^{-3\left(1+w_i\right)h_0}} \nonumber \\
&\, -\frac{\kappa^2}{2}\frac{\frac{9}{2}\sum_{i,j}w_i\left(1+w_i\right)(1+w_j)\rho_{0i}\rho_{0j} a_0^{-3(2+w_i+w_j)} \e^{-3(2+w_i+w_j)h_0}}{\frac {3}{\kappa^2}
\ddot h_0 + \frac{3}{2} \sum _i \left(1+w_i\right) \rho _{0i} a_0^{-3\left(1+w_i\right)} \e^{-3\left(1+w_i\right)h_0}}\, .
\end{align}
Although the expressions in \eqref{M3_4}, \eqref{M3_5} and \eqref{M3_6} are quite complicated, we can still check stability in principle.

\subsection{More General Models}

Since in the model \eqref{M1}, the transition between the non-phantom Universe and the phantom Universe cannot be realized, we generalize the model as follows,
\begin{align}
\label{KK1}
S=  \int d^4 x \sqrt{-g} \left( \frac{R}{2\kappa^2} - K \left( \phi, X \right) + L_\mathrm{matter}\right)\, ,\quad
X \equiv \partial^\mu \phi \partial_\mu \phi \, .
\end{align}
The corresponding FLRW equations have the following forms,
\begin{align}
\label{KK2}
\frac{3}{\kappa^2} H^2 = 2 X \frac{\partial K\left(\phi, X \right)}{\partial X} - K\left( \phi, X \right) + \rho_\mathrm{matter}\, ,\quad
 - \frac{1}{\kappa^2}\left(2 \dot H + 3 H^2 \right) = K\left( \phi, X \right) + p_\mathrm{matter}\left(t\right)\, .
\end{align}
The equation given by the variation of the action~\eqref{KK1} with
respect to $\phi$ is obtained from the equations in \eqref{KK2}
when we use the conservation law of matter,
$0=\dot\rho_\mathrm{matter} + 3H
\left(p_\mathrm{matter}+\rho_\mathrm{matter}\right)$. Therefore,
we can forget the equation.  As in \eqref{M15}, we consider the
matter fluids which have constant EoS parameters $w_i$. By
choosing $\phi=t$, again, we rewrite the equations in \eqref{KK2}
as follows,
\begin{align}
\label{KK4}
K\left( t, -1 \right) = - \frac{1}{\kappa^2}\left(2 \dot H + 3 H^2 \right)  - \sum_i w_i\rho_{0i} a^{-3\left(1+w_i\right)}\, ,\quad
\left. \frac{\partial K\left( \phi, X \right)}{\partial X}\right|_{X=-1} =  \frac{1}{\kappa^2} \dot H  + \frac{1}{2}\sum_i \left(1+w_i\right) \rho_{0i} a^{-3\left(1+w_i\right)}\, .
\end{align}
By using a given function $h\left(\phi\right)$ and choosing $K(\phi,X)$ as follows,
\begin{align}
\label{KK5}
K(\phi,X) =&\, \sum_{n=0}^\infty \left(X+1\right)^n K^{(n)} \left(\phi\right) \, , \nonumber \\
K^{(0)} \left( \phi \right) =&\, - \frac{1}{\kappa^2}\left(2 h''\left(\phi\right) + 3 h'\left(\phi\right)^2 \right)
 - \sum_i w_i\rho_{0i} a_0^{-3\left(1+w_i\right)}\e^{-3\left(1+w_i\right)h\left(\phi\right)} \, , \nonumber \\
K^{(1)} \left( \phi \right) =&\, \left\{  \frac{1}{\kappa^2}
h''\left(\phi\right)  + \frac{1}{2}\sum_i \left(1+w_i\right) \rho_{0i}
a_0^{-3\left(1+w_i\right)}\e^{-3\left(1+w_i\right)h\left(\phi\right)}\right\} \, ,
\end{align}
we find that a solution of the FLRW equations in \eqref{KK2} is given by,
\begin{align}
\label{KK6} H= h'\left(t\right) \quad \left(a = a_0
\e^{h\left(t\right)} \right)\, , \quad \phi=t \, .
\end{align}
We should note $K^{(n)}\left(\phi\right)$ with $n=2,3,\cdots$ can
be arbitrary functions in \eqref{KK5}. The case that
$K^{(n)}\left(\phi\right)$'s with $n=2,3,\cdots$ vanish was
studied in \cite{Nojiri:2005pu, Capozziello:2005tf} and the
instability was investigated. It is straightworward that
\eqref{KK6} is a solution of the equations \eqref{KK2} because the
equations in \eqref{KK4} are obtained.

We now investigate the stability of the equations without matter.
 From the equations in \eqref{KK2}, we can derive the following equation, which does not contain the variable $h''$,
\begin{align}
\label{KK7}
3 \frac{1-y^2}{1+X}X= -\frac{\dot H}{H^2} + \frac{\kappa^2}{H^2} \sum^{\infty}_{n=2} \left( (n-1)X-n-1 \right)X(X+1)^{n-2}K^{(2)}\left(\phi\right)\, ,
\end{align}
where $y=\frac{h'}{H}$. Using Eq.~\eqref{KK7}, we can rewrite
$dy/dN = \left(1/H\right)dy/dt$ ($N$ is called e-foldings and the
scale factor is given in terms of $N$ as $a\propto \e^N$) in the
form which does not contain $h(t)$,
\begin{align}
\label{KK8}
\frac{dy}{dN} = 3X \frac{1-y^2}{1+X} \left(\frac{\dot \phi}{X} +y \right)
 - \frac{\kappa^2}{H^2} \sum^{\infty}_{n=2} \left[ ( \dot \phi +yX) \left( (n-1)X -n-1 \right)
+ \dot \phi n(X+1) \right](X+1)^{n-2}K^{(2)}\left(\phi\right)\, .
\end{align}
We now consider the perturbation from a solution $\phi=t$ in \eqref{KK8} by putting,
\begin{align}
\label{prtph}
\phi = t + \delta\phi \, .
\end{align}
First, we should note that in the limit of the background
solution, we have $X\to -1$ and $y\to 1$. In the limit, the r.h.s.
of \eqref{KK7} is finite and therefore $\frac{1-y^2}{1+X}$ is
finite in the limit,
\begin{align}
\label{yX}
\frac{1-y^2}{1+X} \to \frac{1-y}{1-\dot\phi} = - \frac{\dot H}{3H^2} - \frac{2\kappa^2}{3H^2} K^{(n)}\, .
\end{align}
Here we use $X=-{\dot\phi}^2$.
Then we find,
\begin{align}
\label{phiX}
\frac{\dot \phi}{X} + y
\to \left( 1 - \dot\phi \right) \left( 1 - \frac{\dot H}{3H^2} - \frac{2\kappa^2}{3H^2} K^{(n)} \right) \, .
\end{align}
Eq.~\eqref{yX} also tells
\begin{align}
\label{yphi}
\delta y = \left( 1 - \frac{\dot H}{3H^2} - \frac{2\kappa^2}{3H^2} K^{(2)} \right) \delta\dot\phi
\to  \left( 1 - \frac{h''}{3{h'}^2} - \frac{2\kappa^2}{3{h'}^2} K^{(2)} \right) \delta\dot\phi \, ,
\end{align}
and also, we find
\begin{align}
\label{dya}
\delta y \to \frac{h''}{h'}\delta\phi - \frac{h''}{h'} \delta H \, .
\end{align}
By combining \eqref{yphi} and \eqref{dya}, the following expression of $\delta H$ is obtained,
\begin{align}
\label{deltaH}
\delta H = \delta \phi +\frac{h'}{h''} \left( 1 - \frac{h''}{3{h'}^2} - \frac{2\kappa^2}{3{h'}^2} K^{(2)} \right) \delta\dot\phi
\end{align}
Then we find
\begin{align}
\label{delta}
\frac{d \delta\dot\phi}{dN}
=&\, \left[ - \left( 1 - \frac{h''}{3{h'}^2} - \frac{2\kappa^2}{3{h'}^2} K^{(2)} \right)^{-1} \left\{- 3+ \frac{d}{dN} \left( - \frac{h''}{3{h'}^2} - \frac{2\kappa^2}{3{h'}^2} K^{(2)} \right)\right\}
+ \left( - \frac{h''}{{h'}^2} - \frac{2\kappa^2}{{h'}^2} K^{(2)} \right) \right] \delta\dot\phi
\end{align}
if the quantity inside the square brackets in \eqref{delta} is
negative, the fluctuation $\delta\dot\phi$ becomes exponentially
smaller with time, and therefore the solution becomes stable. Note
that the stability is determined only in terms of $K^{(2)}$ and
does not depend on other $K^{(n)}$ ($n\neq 2$). Then if we choose
$K^{(2)}$ properly, the solution corresponding to an arbitrary
evolution of the Universe becomes stable. As shown in
\cite{Matsumoto:2010uv}, $K^{(2)}$ plays an important role in the
stability, although $K^{(2)}$ does not contribute to the evolution
of the FLRW universe.

We now investigate the stability when we include the matter.
The equation~\eqref{KK7} is modified as follows,
\begin{align}
\label{KK18}
3 \frac{1-y^2}{1+X}X =&\, -\frac{\dot H}{H^2} + \frac{\kappa^2}{H^2} \sum^{\infty}_{n=2} \left( (n-1)X-n-1\bigg)X(X+1)^{n-2}K^{(n)}\left(\phi\right) \right. \nonumber \\
&\, + \frac{\kappa^2}{H^2} \frac{X-1}{2(X+1)} \rho_\mathrm{matter}
 - \frac{\kappa^2}{2H^2}p_\mathrm{matter}-\frac{\kappa^2}{H^2}\frac{X}{X+1} \sum_i \rho _{0i} a_0^{-3\left(1+w_i\right)} \e^{-3\left(1+w_i\right)h\left(\phi\right)}\, .
\end{align}
Then we obtain,
\begin{align}
\label{KK19}
\frac{dy}{dN} =&\, 3X \frac{1-y^2}{1+X} \left(\frac{\dot \phi}{X} +y \right)
 - \frac{\kappa^2}{H^2} \sum^{\infty}_{n=2} \left[ ( \dot \phi +yX) \left( (n-1)X -n-1 \right)
+ \dot \phi n(X+1) \right](X+1)^{n-2}K^{(n)}\left(\phi\right) \nonumber \\
&\, + \frac{\kappa^2}{2H^2 X} \left( -\frac{X-1}{X+1}\left( \dot \phi + yX \right) - \dot \phi \right)
\rho_\mathrm{matter} + \frac{\kappa^2}{2H^2}yp_\mathrm{matter} \nonumber \\
&\, + \frac{\kappa^2}{2H^2} \sum_i \left( \left( \dot \phi +yX \right) \frac{2}{X+1} - \dot \phi \left(1+w_i\right) \right) \rho_{0i} a_0^{-3\left(1+w_i\right)}
\e^{-3\left(1+w_i\right)h\left(\phi\right)}\, .
\end{align}
The inclusion of matter makes the situation complicated because not only $H$ but the scale factor $a$ appears in the equation.
Therefore, we need to include the equation, which describes the evolution of $a$. By defining $\delta\lambda$ by,
\begin{align}
\label{KK24}
\delta \lambda \equiv 3 \sum _i \left(1+w_i\right)
\rho _{0i} {a_0}^{-3\left(1+w_i\right)} \e^{-3\left(1+w_i\right)h}
\frac{\kappa^2}{6 {h'}^2} \left( \frac{\delta a}{a} -g ' \delta \phi \right)\, ,
\end{align}
we obtain the following equations,
\begin{align}
\label{KK29}
\left. \frac{d}{dN} \left(
\begin{array}{c}
 \delta \dot \phi \\
 \delta \lambda
\end{array}
\right) \right| _{\phi =t, H=h'\left(t\right)} = \left(
\begin{array}{cc}
A & B \\
C & D
\end{array}
\right) \left(
\begin{array}{c}
 \delta \dot \phi \\
 \delta \lambda
\end{array}
\right) \, .
\end{align}
Here,
\begin{align}
\label{KK30}
A \equiv&\, -3 + \frac{h''}{{h'}^2} + \frac{\kappa^2}{2 {h'}^2}
\sum _i \left(1+w_i\right) \rho _{0i} {a_0}^{-3\left(1+w_i\right)} \e^{-3\left(1+w_i\right)h} \nonumber \\
&\, - \frac{d}{dN} \ln \left\{ 8K^{(2)} - \frac{2}{\kappa^2}h'' - \sum _i \left(1+w_i\right) \rho _{0i} {a_0}^{-3\left(1+w_i\right)} \e^{-3\left(1+w_i\right)h} \right\} \, ,\nonumber \\
B \equiv&\, 3 - \frac{24 K^{(2)}}{8K^{(2)} - \frac{2}{\kappa^2} h'' - \sum _i \left(1+w_i\right) \rho _{0i} {a_0}^{-3\left(1+w_i\right)} \e^{-3\left(1+w_i\right)h}} \, , \nonumber \\
C \equiv&\, 3 \sum _i \left(1+w_i\right) \rho _{0i} {a_0}^{-3\left(1+w_i\right)} \e^{-3\left(1+w_i\right)h} \left( \frac{\kappa^2}{6 {h'}^2} \right)^2 \nonumber \\
&\, \times \left\{ 8K^{(2)}- \frac{6{h'}^2}{\kappa^2} - \frac{2h''}{\kappa^2}
 - \sum _i \left(1+w_i\right) \rho _{0i} {a_0}^{-3\left(1+w_i\right)} \e^{-3\left(1+w_i\right)h} \right\} \, , \nonumber \\
D \equiv&\,  \frac{d}{dN} \ln \left\{ \frac{\kappa^2}{2 {h'}^2}
\sum _i \left(1+w_i\right) \rho _{0i} {a_0}^{-3\left(1+w_i\right)} \e^{-3\left(1+w_i\right)h} \right\}
 - \frac{\kappa^2}{2{h'}^2} \sum _i \left(1+w_i\right) \rho _{0i} {a_0}^{-3\left(1+w_i\right)} \e^{-3\left(1+w_i\right)h} \, .
\end{align}
Generally, the $2\times 2$ matrix must have a negative trace and a positive determinant so that the two eigenvalues of the matrix could be negative because the two eigenvalues are given by $\frac{1}{2}\{ \mathrm{tr} M \pm \sqrt{(\mathrm{tr} M)^2-4(\det M)} \}$ for $M\equiv \left( \begin{array}{cc} A & B \\ C & D \end{array} \right)$.
In order to investigate the stability of the fixed point $\phi =t$, $H=h'\left(t\right)$, we only need to calculate the determinant and the trace of the matrix in Eq.~\eqref{KK29} with \eqref{KK30}.

The condition of stability is given by the inequality, which tells that there
%is
could be
%a wide class of
stable models.

\section{Eliminating Ghosts and Phantom Crossing}

\subsection{Absence of Ghosts}

Although it was not explicitly written in Ref.~\cite{Matsumoto:2010uv}, we show that we can exclude ghosts even in the phantom Universe in the model \eqref{KK1}.
To this purpose, we consider the perturbation of $\phi$ from the background solution $\phi=t$ as in Eq.~\eqref{prtph}.
When we substitute the expression into $K(\phi,X)$ in \eqref{KK5}, if the coefficient of $\delta{\dot\phi}^2$ is positive, the ghosts do not appear.
By the substitution, we obtain,
\begin{align}
\label{KK5prt}
K(\phi,X) =&\, \left. \left\{ - K^{(1)} \left( \phi \right) + 8 K^{(2)} \left(\phi\right) \right\} \right|_{\phi=t} \delta{\dot\phi}^2 \nonumber \\
&\, + \left( \mbox{terms which do not include $\delta{\dot\phi}^2$} \right) + \mathcal{O}\left({\delta\phi}^3\right) \, .
\end{align}
Therefore, if we choose $K^{(2)} \left(\phi\right)$ by using a parameter $\mu$ with the dimension of mass, for example,
\begin{align}
\label{prtph2}
K^{(2)} \left(\phi\right) = \frac{1}{8} \left\{ \frac{1}{2\mu^4} + K^{(1)} \left( \phi \right) \right\}\, ,
\end{align}
ghosts do not appear because,
\begin{align}
\label{KK5prt2}
K(\phi,X) =&\, \frac{1}{2\mu^4} \delta{\dot\phi}^2
+ \left( \mbox{terms which do not include $\delta{\dot\phi}^2$} \right) + \mathcal{O}\left({\delta\phi}^3\right) \, .
\end{align}
Therefore, $K^{(2)} \left(\phi\right)$ is important not only for
the stability but for the absence of ghosts. The
condition~\eqref{prtph2} includes a parameter $\mu$, and the
condition can be easily relaxed by using more general function
instead of $\mu$, which tells that there is a wide class of
ghost-free models.
 Note that the $k$-essence model above is not fine-tuned. The functions are general and are not specifically chosen to remove ghosts and
to produce the desired crossing. There is a large number of
functions that ensure the stability and fit the desired
description of phantom crossing.

\subsection{Phantom crossing without Ghost Fields}

Since we found the condition that enables us to exclude the ghost degrees of freedom, we consider the model that generates the phantom-to-quintessence transition, as suggested by the DESI observations.

We assume that the energy density $\rho_\mathrm{DE}$ and the pressure $p_\mathrm{DE}$ of the dark energy satisfy the conservation law,
\begin{align}
\label{SGBEG9DE}
0 = \dot \rho_\mathrm{DE} + 3H \left( \rho_\mathrm{DE} + p_\mathrm{DE} \right) \, .
\end{align}
On the phantom crossing,  because $\rho_\mathrm{DE}=-p_\mathrm{DE}$, we find $\dot\rho_\mathrm{DE}$ must vanish.
Because $\rho_\mathrm{DE}<-p_\mathrm{DE}$ in the phantom Universe and $\rho_\mathrm{G}>p_\mathrm{G}$ in the non-phantom Universe, if $\ddot\rho_\mathrm{DE}>0$ when $\rho_\mathrm{DE}=-p_\mathrm{DE}$, a crossing from the non-phantom Universe to the phantom one occurs.
On the other hand, if $\ddot\rho_\mathrm{DE}<0$ when $\rho_\mathrm{DE}=-p_\mathrm{DE}$, the inverse phantom crossing, that is, the transition from the phantom Universe to the non-phantom one, as observed in the DESI, occurs.

For simplicity, as a matter fluid we may include only pressureless dust, which could be cold dark matter and baryonic matter,
\begin{align}
\label{DST}
\rho_\mathrm{dust} = \rho_{0\, \mathrm{dust}} a(t)^{-3} \, ,
\end{align}
where $\rho_{0\, \mathrm{dust}}$ is a positive constant.
In the case of the $\Lambda$CDM model, which is Einstein's gravity coupled with the dust and a cosmological term with a cosmological constant $\Lambda$, the first FLRW equation is given by,
\begin{align}
\label{LCDM}
\frac{3{h'}^2}{\kappa^2} = \frac{\Lambda}{\kappa^2} + \rho_{0\, \mathrm{dust}} {a_0}^{-3} \e^{-3h} \, ,
\end{align}
the solution of $a=a_0\e^{h(t)}$ is given by,
\begin{align}
\label{LCDMsln}
h(t) = \frac{2}{3} \ln \left( \sinh \left(\alpha t\right) \right) + \beta \, ,
\end{align}
with constants $\alpha$ and $\beta$, which are defined as
\begin{align}
\label{alphac}
\alpha \equiv \frac{1}{2}\sqrt{3\Lambda} \, , \quad
\beta \equiv \frac{1}{3} \ln \frac{\kappa^2 \rho_{0\, \mathrm{dust}}}{\Lambda {a_0}^3}\, .
\end{align}
We now deform the behavior of the $\Lambda$CDM model in \eqref{LCDMsln} to include the phantom crossing.
In the first FLRW equation in \eqref{LCDM}, we add the contribution $\rho_k$ from $k$-essence,
\begin{align}
\label{LCDMk}
\frac{3{h'}^2}{\kappa^2} = \rho_k + \frac{4\alpha^2}{3\kappa^2} \left( 1 + \e^{-3h+3\beta} \right)\, ,
\end{align}
where we used \eqref{LCDMsln}.
The contribution from the cosmological constant $\frac{\Lambda}{\kappa^2} = \frac{4\alpha^2}{3\kappa^2}$ may also come from the $k$-essence model, but for convenience, we separate it from the contribution from the $k$-essence model.
Then we may identify,
\begin{align}
\label{Lmdk}
\rho_\mathrm{DE} = \rho_k + \frac{4\alpha^2}{3\kappa^2} \, ,
\end{align}
and we consider $\rho_\mathrm{DE}$ or $\rho_k$, which generates the phantom crossing.
We modify the behavior of $h(t)$ by adding a function $\gamma(t)$ as follows,
\begin{align}
\label{LCDMslnGmm}
h(t) = \frac{2}{3} \ln \left( \sinh \left(\alpha t\right) \right) + \beta + \frac{2}{3}\gamma(t)\, .
\end{align}
Then Eq.~\eqref{LCDMk} gives,
\begin{align}
\label{rhk0}
\rho_k (t)= \frac{4}{3\kappa^2} \left\{ \frac{\alpha^2}{\sinh^2 \left( \alpha t \right)} \left( 1 - \e^{-2\gamma(t)} \right) + 2 \alpha \coth \left( \alpha t \right) \gamma'(t) + \gamma'(t)^2 \right\} \, .
\end{align}
Because,
\begin{align}
\label{rhk1}
{\rho_k}' (t)
%=&\, \frac{4}{3\kappa^2} \left\{ - \frac{2\alpha^3\cosh \left( \alpha t \right)}{\sinh^3 \left( \alpha t \right)} \left( 1 - \e^{-2\gamma(t)} \right)
%+ \frac{2\alpha^2 \gamma'(t) \e^{-2\gamma(t)} }{\sinh^2 \left( \alpha t \right)} \right. \nonumber \\
%&\, \left. - \frac{2 \alpha^2}{\sinh^2 \left( \alpha t \right)} \gamma'(t) + 2 \alpha \coth \left( \alpha t \right) \gamma''(t)
%+ 2 \gamma'(t) \gamma''(t) \right\} \, , \nonumber \\
=&\, \frac{4}{3\kappa^2} \left\{ - \left( \frac{2\alpha^3\cosh
\left( \alpha t \right)}{\sinh^3 \left( \alpha t \right)} + \frac{2 \alpha^2}{\sinh^2 \left( \alpha t \right)} \gamma'(t) \right) \left( 1 - \e^{-2\gamma(t)} \right)
+ 2 \left( \alpha \coth \left( \alpha t \right) + \gamma'(t) \right) \gamma''(t) \right\} \, , \\
\label{rhk2}
{\rho_k}''(t)
%=&\, \frac{4}{3\kappa^2} \left\{ - \left( \frac{2\alpha^4}{\sinh^2 \left( \alpha t \right)}
% - \frac{6\alpha^4\cosh^2 \left( \alpha t \right)}{\sinh^4 \left( \alpha t \right)}
% - \frac{2 \alpha^3 \cosh \left( \alpha t \right)}{\sinh^3 \left( \alpha t \right)} \gamma'(t)
%+  \frac{2 \alpha^2}{\sinh^2 \left( \alpha t \right)} \gamma''(t)
% \right) \left( 1 - \e^{-2\gamma(t)} \right) \right. \nonumber \\
% &\, \left. - 2 \left( \frac{2\alpha^3\cosh \left( \alpha t \right)}{\sinh^3 \left( \alpha t \right)}
% + \frac{2 \alpha^2}{\sinh^2 \left( \alpha t \right)} \gamma'(t) \right) \gamma' \e^{-2\gamma(t)}
% + 2 \left( \frac{\alpha^2}{\sinh^2 \left( \alpha t \right)} + \gamma''(t) \right) \gamma''(t)
% + 2 \left( \alpha \coth \left( \alpha t \right) + \gamma'(t) \right) \gamma'''(t) \right\} \, , \nonumber \\
=&\, \frac{4}{3\kappa^2} \left\{ - \left( - \frac{4\alpha^4}{\sinh^2 \left( \alpha t \right)}
 - \frac{6\alpha^4}{\sinh^4 \left( \alpha t \right)}
 - \frac{2 \alpha^3 \cosh \left( \alpha t \right)}{\sinh^3 \left( \alpha t \right)} \gamma'(t)
+  \frac{2 \alpha^2}{\sinh^2 \left( \alpha t \right)} \gamma''(t) \right) \left( 1 - \e^{-2\gamma(t)} \right) \right. \nonumber \\
&\, - 2 \left( \frac{2\alpha^3\cosh \left( \alpha t \right)}{\sinh^3 \left( \alpha t \right)}
+ \frac{2 \alpha^2}{\sinh^2 \left( \alpha t \right)} \gamma'(t) \right) \gamma' \e^{-2\gamma(t)} \nonumber \\
&\, \left. + 2 \left( \frac{\alpha^2}{\sinh \left( \alpha t \right)} + \gamma''(t) \right) \gamma''(t)
+ 2 \left( \alpha \coth \left( \alpha t \right) + \gamma'(t) \right) \gamma'''(t) \right\} \, , \nonumber \\
\end{align}
We assume ${\rho_k}' (t_0)=0$ for a fixed time $t_0$, when the phantom crossing occurs.
Just for simplicity, we may also assume $\gamma'(t_0) = \gamma'''(t_0) = 0$.
Then Eqs.~\eqref{rhk1} and \eqref{rhk2} gives,
\begin{align}
\label{rhk1B}
0 =&\, \frac{4}{3\kappa^2} \left\{ - \frac{2\alpha^3\cosh \left( \alpha t_0 \right)}{\sinh^3 \left(
\alpha t_0 \right)} \left( 1 - \e^{-2\gamma(t_0)} \right)
+ 2 \alpha \coth \left( \alpha t_0 \right) \right\} \gamma''(t_0)\, , \\
\label{rhk2B}
{\rho_k}''(t_0) =&\, \frac{4}{3\kappa^2} \left\{ - \left( - \frac{4\alpha^4}{\sinh^2 \left( \alpha t_0 \right)}
 - \frac{6\alpha^4}{\sinh^4 \left( \alpha t_0 \right)}
+  \frac{2 \alpha^2}{\sinh^2 \left( \alpha t_0 \right)} \gamma''(t_0)
\right) \left( 1 - \e^{-2\gamma(t_0)} \right) \right. \nonumber \\
&\, \left. + 2 \left( \frac{\alpha^2}{\sinh^2 \left( \alpha t_0 \right)} + \gamma''(t_0) \right) \gamma''(t_0) \right\} \, .
\end{align}
Eq.~\eqref{rhk2B} indicates that,
\begin{align}
\label{rhk3}
\gamma''(t_0) = \frac{\alpha^2}{\sinh^2 \left( \alpha t_0 \right)} \left( 1 - \e^{-2\gamma(t_0)} \right) \, .
\end{align}
By substituting \eqref{rhk3} into \eqref{rhk1B},
\begin{align}
\label{rhk2C}
{\rho_k}''(t_0) =&\, \frac{4}{3\kappa^2} \left( \frac{4\alpha^4}{\sinh^2 \left( \alpha t_0 \right)} + \frac{8\alpha^4}{\sinh^4 \left( \alpha t_0 \right)} \right)
\left( 1 - \e^{-2\gamma(t_0)} \right)  \, .
\end{align}
Then if $\gamma(t_0)>0$, we find ${\rho_k}''(t_0)={\rho_\mathrm{DE}}''>0$, which corresponds to the transition from the non-phantom Universe to the phantom Universe
and if $\gamma(t_0)<0$, we find ${\rho_k}''(t_0)={\rho_\mathrm{DE}}''<0$, that is, the transition from the phantom Universe to the non-phantom Universe.
Eq.~\eqref{rhk3} tells that $\gamma''(t_0)>0$ if $\gamma(t_0)>0$, and $\gamma''(t_0)<0$ if $\gamma(t_0)<0$.
By choosing $\gamma''(t_0)<0$ and $\gamma(t_0)<0$, a simple example is given by,
\begin{align}
\label{ex}
\gamma(t) = \gamma(t_0) - \gamma_0 + \gamma_0
\exp\left( \frac{\gamma''(t_0)}{2\gamma_0} \left(t - t_0\right)^2
\right) \, ,
\end{align}
where $\gamma_0$ is a positive constant.
For the expression \eqref{ex}, it is clear that $\gamma'(t_0) = \gamma'''(t_0) = 0$.
By combining \eqref{LCDMslnGmm} with \eqref{ex}, we find $h(t)$ as a function of $t$.
Then, using \eqref{KK5} with \eqref{prtph2}, we obtain a $k$-essence model that generates the crossing from the phantom to the non-phantom Universe without ghosts.

\subsection{Apparent Phantom Crossing}

In \cite{Khoury:2025txd}, another scenario was proposed that might solve the problem in DESI observations. In the scenario proposed in \cite{Khoury:2025txd}, there is no real transition from $w<-1$ to $w>-1$ in the EoS parameter of dark energy, but the time evolution of dark matter might solve the problem.
The energy density of the dark matter is usually assumed to behave as $a^{-3}$, but if the decrease of the energy density is slower than $a^{-3}$, it looks as though there could be a phantom crossing in the dark energy sector.
In the $k$-essence model \eqref{M1}, the quantization of the scalar field $\phi$ gives massive particles.
Because the massive particles become dust, they might be dark matter.

When we consider the perturbation $\delta\phi$ from the background
$\phi= t$ as in \eqref{prtph}, the quantization of $\delta\phi$
gives the particles corresponding to the dark matter. Let the mass
of the particle corresponding to $\delta\phi$ be $m_{\delta\phi}$.
If the particle is not created or annihilated from ten or more
billion years ago, the number of dark matter particles is
conserved. Then the number density $n_{\delta\phi}$ is
proportional to the inverse of the volume, and we find
$n_{\delta\phi} = n_{\delta\phi\, 0} a^{-3}$ if the particle is
not relativistic. This tells that the energy density of the
particles is given by $\rho_{\delta\phi} =
m_{\delta\phi}n_{\delta\phi} = m_{\delta\phi} n_{\delta\phi\,
0}a^{-3}$. If the mass $m_{\delta\phi}$ increases by the
expansion, the energy density decreases more slowly than $a^{-3}$.

As in \eqref{KK5prt} and \eqref{KK5prt2}, we expand $K(\phi,X)$ by the perturbation \eqref{prtph} and we use \eqref{prtph2},
\begin{align}
\label{KK5fprtF}
K(\phi,X) =&\, \frac{1}{2\mu^4} \delta{\dot\phi}^2 + a(t)^{-2} \left. K^{(1)} \left( \phi \right)
\right|_{\phi=t} \sum_{i=1,2,3} \left( \partial_i \delta\phi \right)^2
 - \left. {K^{(1)}}' \left( \phi \right) \right|_{\phi=t} \delta{\dot\phi} \delta\phi
+ \frac{1}{2} \left. {K^{(0)}}'' \left( \phi \right) \right|_{\phi=t} {\delta\phi}^2
+ \left( \mathcal{O} \left( {\delta\phi}^3 \right) \ \mbox{terms} \right) \nonumber \\
=&\, \frac{1}{2\mu^4} \delta{\dot\phi}^2 + a(t)^{-2} \left.
\left\{ \frac{1}{\kappa^2} h''\left(\phi\right)  + \frac{1}{2}\sum_i \left(1+w_i\right)
\rho_{0i} a_0^{-3\left(1+w_i\right)}\e^{-3\left(1+w_i\right)h\left(\phi\right)} \right\} \right|_{\phi=t} \sum_{i=1,2,3} \left( \partial_i \delta\phi \right)^2 \nonumber \\
&\, - \left. \left\{ \frac{1}{\kappa^2} h'''\left(\phi\right)  - \frac{3}{2}\sum_i \left(1+w_i\right)^2
\rho_{0i} a_0^{-3\left(1+w_i\right)}\e^{-3\left(1+w_i\right)h\left(\phi\right)} h'(\phi) \right\} \right|_{\phi=t} \delta{\dot\phi} \delta\phi \nonumber \\
&\, + \frac{1}{2} \left\{ - \frac{1}{\kappa^2}\left(2 h''''\left(\phi\right) + 6 h''\left(\phi\right)h''\left(\phi\right) + 6 h'\left(\phi\right) h'''\left(\phi\right) \right) \right. \nonumber \\
&\, \left. \left. - \sum_i w_i\rho_{0i} a_0^{-3\left(1+w_i\right)}\e^{-3\left(1+w_i\right)h\left(\phi\right)}
\left( 9\left(1+w_i\right)^2{h'\left(\phi\right)}^2 - 3\left(1+w_i\right)h''\left(\phi\right) \right)
\right\} \right|_{\phi=t} {\delta\phi}^2 \nonumber \\
&\, + \left( \mathcal{O} \left( {\delta\phi}^3 \right) \ \mbox{terms} \right) \, .
\end{align}
The third term on the right-hand side in the first line is rewritten as follows,
\begin{align}
\label{KK5fprtF2}
 - \left. {K^{(1)}}' \left( \phi \right) \right|_{\phi=t} \delta{\dot\phi} \delta\phi
=  - a^{-3} \frac{d}{dt} \left[  \frac{a^3}{2} \left. {K^{(1)}}'
\left( \phi \right) \right|_{\phi=t} {\delta\phi}^2 \right] + \left[ \frac{3H}{2} \left. {K^{(1)}}' \left( \phi \right)
\right|_{\phi=t} + \frac{1}{2} \left. {K^{(1)}}'' \left( \phi \right) \right|_{\phi=t}\right] {\delta\phi}^2 \, .
\end{align}
In the action for the perturbation $S=\int d^4x \sqrt{-g} K(\phi,X) + \cdots = \int d^4x a^3 K(\phi,X) + \cdots$, the second line in \eqref{KK5fprtF2} can be dropped.
We should also note that $H=h'\left(\phi = t \right)$.
By combining \eqref{KK5fprtF} with \eqref{KK5fprtF2}, we obtain,
\begin{align}
\label{KK5fprtF3}
K(\phi,X) =&\, \frac{1}{2\mu^4}
\delta{\dot\phi}^2 + a(t)^{-2} \left. K^{(1)} \left( \phi \right)
\right|_{\phi=t} \sum_{i=1,2,3} \left( \partial_i \delta\phi
\right)^2 + \frac{1}{2} \left[ \left. {K^{(0)}}'' \left( \phi
\right) \right|_{\phi=t}
+ 3H \left. {K^{(1)}}' \left( \phi \right) \right|_{\phi=t} + \left. {K^{(1)}}'' \left( \phi \right) \right|_{\phi=t} \right] {\delta\phi}^2 \nonumber \\
&\, + \left( \mathcal{O} \left( {\delta\phi}^3 \right) \ \mbox{terms and total derivative terms in the action} \right) \nonumber \\
=&\, \frac{1}{2\mu^4} \delta{\dot\phi}^2 + a(t)^{-2} \left.
\left\{ \frac{1}{\kappa^2} h''\left(\phi\right)  + \frac{1}{2}\sum_i \left(1+w_i\right)
\rho_{0i} a_0^{-3\left(1+w_i\right)}\e^{-3\left(1+w_i\right)h\left(\phi\right)} \right\} \right|_{\phi=t} \sum_{i=1,2,3} \left( \partial_i \delta\phi \right)^2 \nonumber \\
&\, - \frac{1}{2} \left[ \frac{1}{\kappa^2} \left\{ h''''\left(\phi\right) + 3 h'\left(\phi\right) h'''\left(\phi\right) + 6 h''\left( \phi \right)^2 \right\} \right. \nonumber \\
&\, \left. \left. + \frac{1}{2} \sum_i \rho_{0i} a_0^{-3\left(1+w_i\right)}\e^{-3\left(1+w_i\right)h\left(\phi\right)}
\left\{ 9 w_i \left(1+w_i\right)^2 h'\left(\phi\right)^2 + 3 \left(1 - {w_i}^2\right) h''(\phi) \right\} \right] \right|_{\phi=t} {\delta\phi}^2 \nonumber \\
&\, + \left( \mathcal{O} \left( {\delta\phi}^3 \right) \ \mbox{terms and total derivative terms in the action} \right) \, .
\end{align}
Then the mass $m_{\delta\phi}$ of the particle corresponding to $\delta\phi$ is identified with,
\begin{align}
\label{mdltph}
{m_{\delta\phi}}^2 =&\, \mu^4 \left[ \left.
{K^{(0)}}'' \left( \phi \right) \right|_{\phi=t}
+ 3H \left. {K^{(1)}}' \left( \phi \right) \right|_{\phi=t} + \left. {K^{(1)}}'' \left( \phi \right) \right|_{\phi=t} \right] \nonumber \\
=&\, \mu^4 \left[ \frac{1}{\kappa^2} \left\{ h''''\left(\phi\right) + 3 h'\left(\phi\right) h'''\left(\phi\right) + 6 h''\left( \phi \right)^2 \right\} \right. \nonumber \\
&\, \left. \left. + \frac{1}{2} \sum_i \rho_{0i}
a_0^{-3\left(1+w_i\right)}\e^{-3\left(1+w_i\right)h\left(\phi\right)}
\left\{ 9 w_i \left(1+w_i\right)^2 h'\left(\phi\right)^2 + 3 \left(1 - {w_i}^2\right) h''(\phi)
\right\}  \right] \right|_{\phi=t} \nonumber \\
=&\, \frac{\mu^4}{\kappa^2} \left\{ \dddot H + 3 H \ddot H + 6 {\dot H}^2 \right\} + \frac{\mu^4}{2} \sum_i \rho_{0i}
a(t)^{-3\left(1+w_i\right)} \left\{ 9 w_i \left(1+w_i\right)^2 H^2
+ 3 \left(1 - {w_i}^2\right) \dot H \right\} \, .
\end{align}
This tells that the mass $m_{\delta\phi}$ changes due to the expansion of the Universe. In order to avoid the tachyon instability, we need to require ${m_{\delta\phi}}^2\geq 0$.
If the mass $m_{\delta\phi}$ increases, the apparent phantom crossing could occur.

For simplicity, we only consider the dust with $w_i=0$ as matter.
The expression \eqref{mdltph} tells that,
\begin{align}
\label{mdltph2}
{m_{\delta\phi}}^2 =&\, \frac{1}{\kappa^2} \left\{
\dddot H + 3 H \ddot H + 6 {\dot H}^2 \right\} + \frac{3}{2} \rho_{\mathrm{dust}\,0} a(t)^{-3}\dot H \, ,
\end{align}
and
\begin{align}
\label{mdltph3}
\frac{d{m_{\delta\phi}}^2}{dt} =&\,
\frac{1}{\kappa^2} \left\{ \ddddot H + 3 H \dddot H + 15 \dot H
\ddot H \right\} + \frac{3}{2} \rho_{\mathrm{dust}\,0} \frac{d \left(a(t)^{-3}\dot H\right)}{dt} \, .
\end{align}
If the total energy and the total pressure satisfy the energy conditions, $\dot H\leq 0$.
Furthermore, suppose we assume that the Universe is asymptotically de Sitter space.
In that case, $H$ goes to a constant, or if we assume that the Universe is an asymptotically power-law expanding universe, $H$ goes to proportional to $t^{-1}$.
For the assumptions,  $\dot H$ and $a(t)^{-3}\dot H$ must be a negative and asymptotically increasing function.
Furthermore, $\ddot H$,  $\ddddot H$ and $\frac{d \left(a(t)^{-3}\dot H\right)}{dt}$ are positive and asymptotically decreasing functions, and $\dddot H$ is a negative and asymptotically increasing function.
Therefore, if $\ddddot H$ or $\frac{d \left(a(t)^{-3}\dot H\right)}{dt}$ dominates in \eqref{mdltph3}, $m_{\delta\phi}$ is an asymptotically increasing function, and it may generate the apparent (inverse) phantom crossing.

It is not so straightforward to realize the (inverse) phantom crossing.
As an example, we consider the matter-dominated Universe, $a \propto t^\frac{2}{3}$ and $\frac{3H^2}{\kappa^2} \sim \rho_{\mathrm{dust}\,0} a(t)^{-3}$ which was realized before the accelerating expansion of the Universe started.
Then, since $H\sim \frac{2}{3t}$, $\rho_{\mathrm{dust}\,0} a(t)^{-3} \sim \frac{4}{3\kappa^2 t^2}$, $\dot H \sim - \frac{2}{3t^2}$, $\ddot H \sim \frac{4}{3t^3}$, $\dddot H \sim - \frac{4}{t^4}$ and
$\ddddot H \sim \frac{16}{t^5}$,  we find,
\begin{align}
\label{mdltph4}
{m_{\delta\phi}}^2 \sim &\, \frac{1}{\kappa^2 t^4}
\left( - 4 - \frac{8}{3} + \frac{8}{3} - \frac{4}{3} \right) = - \frac{16}{3 \kappa^2 t^4} \\, ,
\end{align}
Then the particle corresponding to $\delta\phi$ becomes a tachyon.
Here, we neglected the contribution from the $k$-essence itself in the background evolution and the back-reaction from the $\delta\phi$ particle itself.
The situation might be improved by a more detailed analysis.

\section{Connecting Inflation to the Dark Energy Epoch: Some Qualitative Considerations}

One of the demanding tasks that a modern theoretical physicist is
required to answer concretely is what is the driving force behind
the inflationary era and the dark energy era. Do these quasi-de
Sitter epochs have the same source, and if yes, how is it possible
to discriminate among theoretical models that can unify inflation
with dark energy? DESI provided hints that point out towards
modified gravity, since phantom-to-quintessence transitions are
possible in the context of GR, but only in $k$-essence, canonical
single field models are excluded. But the DESI data do not provide
a decisive hit against single scalar field theory. Now, in this
section, we shall consider the distinctive character of the
inflationary and dark energy quasi-de Sitter epochs in single
scalar field theory and in one popular modified gravity, namely
$F(R)$ gravity. The EoS parameter of the two theories is
significantly constrained from the updated Planck/BICEP data,
which constrain the tensor-to-scalar ratio as
\cite{BICEP:2021xfz},
\begin{align}
\label{planck}
r<0.036\, .
\end{align}
Now we shall compare the EoS parameter for inflationary theories in the context of minimally coupled canonical single scalar field theory and of $F(R)$ gravity. In the former, the EoS parameter is,
\begin{align}
\label{eossingle}
w=-1+\frac{r}{24}\, ,
\end{align}
while in the case of $F(R)$ gravity, the EoS parameter is,
\begin{align}
\label{freos}
w=-1+\frac{2r^{1/2}}{3\sqrt{48}}\, .
\end{align}
The above relation (\ref{freos}) is obtained easily from the
general relation in every FRW cosmology
\begin{equation}\label{ref1fort}
w=-1-\frac{2}{3}\frac{\dot{H}}{H^2}\, ,
\end{equation}
which is expressed in terms of the first slow-roll index
$\epsilon_1=-\frac{\dot{H}}{H^2}$, so we have,
\begin{equation}\label{ref2fort}
w=-1+\frac{2}{3}\epsilon_1\, .
\end{equation}
Now since for $F(R)$ gravity we have $r=48\epsilon_1^2$
\cite{Oikonomou:2025qub}, substituting $r=48\epsilon_1^2$ in Eq.
(\ref{ref2fort}) one easily obtains Eq. (\ref{freos}). Thus, since
$r<0.036$, this means that for scalar theories, the EoS parameter
is,
\begin{align}
\label{rscalar}
w_\mathrm{scalar}<-0.9985\, ,
\end{align}
while for $F(R)$ gravity, the EoS parameter must be,
\begin{align}
\label{eoesfrcons}
w_{F(R)}<-0.9817425\, .
\end{align}
Going to the dark energy era, the quasi-de Sitter epoch EoS
parameter is constrained to be $w=-0.997\pm 0.025$
\cite{DESI:2025zgx}, note however this constraint is obtained by
using a constant EoS, but we use it as a benchmark value. Thus,
this result does not say much about the generating theory of
inflation and dark energy, and it also does not say anything on
whether the generating theory behind inflation is the same as the
theory behind dark energy. The slight difference between $F(R)$
and single scalar theory is that the EoS parameter in $F(R)$
gravity is allowed to take slightly but measurable values compared
to the single scalar field description. Hence, if the future stage
4 Cosmic Microwave Background (CMB) experiments reveal a
tensor-to-scalar ratio of the order $\sim 0.03$, and dark energy
experiments point towards larger values of the EoS parameter, the
result might indicate that modified gravity might be the common
source of the two quasi-de Sitter epochs. Regarding the
flexibility of GR to describe both inflation and dark energy, the
theory is very much constrained in order to also accommodate the
phantom-to-quintessence transition. Indeed, the unified theory
must describe inflation in a viable way, and dark energy in a
viable way, but we did not pursue our analysis to this level, this
can be done in a future article focused on exactly that. Now, one
thing is certain: single scalar field theory cannot be the source
of inflation and dark energy and at the same time provide a
phantom-to-quintessence transition.

\section{Conclusions}

Although GR is well-proven and established at astrophysical
levels, it seems that at large scales, it has serious issues that
demand a compelling explanation. Specifically, the latest DESI
data have motivated renewed interest in dynamical dark energy and
possible departures from a cosmological constant with a preference
for phantom crossing in the dark energy sector.
%DESI data shook the ground for GR descriptions significantly since the data point out towards a dynamical dark energy, but more importantly, they seem to favor a phantom-to-quintessence transition in the late Universe.
In this work, we concretely investigated whether a
phantom-to-quintessence can occur in the context of GR. The
phantom-to-quintessence transition cannot occur for sure in the
context of single scalar field theory, even with negative
potentials being used. However, the $k$-essence theories are still
theories harbored in the GR context. These $k$-essence models can,
in principle, realize a phantom-to-quintessence transition, and
the detailed theory behind this transition was the focus of this
work. We answered why a phantom-to-quintessence transition cannot
be realized by a canonical and minimally coupled single scalar
field in great detail, even if negative potentials are used. Also,
for the $k$-essence theories, we discussed how ghost instabilities
might occur, how the ghosts could be eliminated, and how such a
phantom-to-quintessence transition might be eventually realized by
$k$-essence theories. However, simply realizing the
phantom-to-quintessence transition does not make these theories
viable. One must produce an observationally viable dark energy
era, compatible with all the observational data, and also must
explain whether inflation and dark energy can be described by the
same $k$-essence theory. This is not an easy task for
 $k$-essence models, although the unified and viable
description of inflation and dark energy is a matter of choosing a
model in $F(R)$ gravity. To this end, we qualitatively discussed
how modified gravity might manifest itself in the dark energy
data, using also CMB constraints. As we indicated, a more
quintessential EoS parameter might be a hint of modified gravity.
Note that $k$-essence models and minimally coupled quintessence
models belong to the framework of GR, while extensions of GR
include modifications of the Einstein-Hilbert Lagrangian, like
$F(R)$ gravity, or even terms in the Lagrangian that come from
one-loop corrections of the scalar field Lagrangian in its vacuum
configuration (minimally or conformally coupled), so essentially
higher contractions of the Riemann and Ricci tensor and non-local
terms.

Although we clarified the condition of the absence of the ghost,
in order to show the full consistencies, we need to check the
scalar sound speed, possible strong coupling near the crossing,
and the behavior of perturbations through the transition. However,
the primordial scalar perturbations which would involve the sound
wave speed are not the focus of this work, in which we aimed
studying the late-time features of the model, so no scalar or
tensor perturbations are involved. In a unified work the late and
early time behaviors of the model should be studied.

Although we consider the $k$-essence models, there are other
possible models. For example, there are disformal and covariant
dark-sector couplings~\cite{Faraoni:2014vra,Bansal:2025usn,
Khoury:2025txd}, which can produce effective phantom-like
behaviour without introducing a fundamental ghost degree of
freedom, higher-dimensional and braneworld
realizations~\cite{Mishra:2025goj, Pan:2025psn, Wolf:2024stt},
where scalar-field dark energy models on a ghost-free phantom
braneworld can produce effective phantom-divide crossing for
several thawing potentials. Also we should also mention
interacting dark-sector models
\cite{Petri:2025swg,Chakraborty:2025syu,Shah:2024rme} in which
interacting or coupled fluid model can be degenerate with an
evolving dark-energy parametrization at the background level,
while the apparent phantom crossing is replaced by a sign change
in the dark-sector interaction.

We should also note that using the equation of state as the
primary parametrization itself might be partly responsible for
emphasizing phantom crossing. Pressure parameterizations provide
an alternative way to model deviations from a cosmological
constant and can be realized with additional $k$-essence
fields~\cite{Sen:2007gk, Shlivko:2024llw}.

\section*{Statements}

No new data were used for this manuscript.

\end{document}